\documentclass[prl,twocolumn,floatfix,superscriptaddress]{revtex4-1}

\usepackage{graphicx} % enhanced support for graphics
\usepackage{color} % use colored text in latex
\usepackage{amsmath} % Defines extra environments for multiline displayed equations, as well as a number of other enhancements for math
\usepackage[urlcolor=blue, hyperindex, colorlinks, bookmarks=true,linkcolor=black,citecolor=black]{hyperref} % provides LaTeX the ability to create hyperlinks within the document
\usepackage{amssymb}% {bm}% bold math

\newcommand{\bra}[1]{\langle{#1}|}
\newcommand{\ket}[1]{|{#1}\rangle}

\begin{document}

\title{A superconducting qubit with Purcell protection and tunable coupling}
\date{\today}
\author{J. M. Gambetta}
\affiliation{Institute for Quantum Computing and Department of Applied Mathematics, University of Waterloo, Waterloo, Ontario N2L 3G1, Canada}	
\author{A. A. Houck}
\affiliation{Department of Electrical Engineering, Princeton University, Princeton, New Jersey 08544, USA}	
\author{Alexandre Blais}
\affiliation{D\'epartement de Physique, Universit\'e de Sherbrooke, Sherbrooke, Qu\'ebec, Canada, J1K 2R1}

\begin{abstract}
We present a superconducting qubit for the circuit quantum electrodynamics architecture that has a tunable coupling strength $g$. We show that this coupling strength can be tuned from zero to values that are comparable with other superconducting qubits. At $g=0$ the qubit is in a decoherence free subspace with respect to spontaneous emission induced by the Purcell effect. Furthermore we show that in the decoherence free subspace the state of the qubit can still be measured by either a dispersive shift on the resonance frequency of the resonator or by a cycling-type measurement.
\end{abstract}
\maketitle

Quantum decoherence is one of the major problems facing quantum information processing. To overcome this problem the theories of quantum error correction \cite{Knill1997} and decoherence free subspaces (DFS) \cite{Lidar1998} have been developed. A DFS is a subspace of a system which exploits symmetries in the decoherence process to allow the system to be completely decoupled from the environment. As an example, the spontaneous decay of a multilevel atom into the same bath can be cancelled for one of the states by quantum interference \cite{Zhu1996}.

In recent years superconducting qubits have emerged as candidates for quantum information processing \cite{Schoelkopf2008}. These are systems which are designed using Josephson junctions to make low loss non-linear oscillators.  They are designed so that two levels (qubit) can be isolated, controlled and measured, properties which are usually mutually exclusive. With \emph{sweet spot} operations
 \cite{Vion2002,Orlando1999,Koch2007} and material engineering \cite{Martinis2005} there has been tremendous progress.
 This is evidenced by the recent demonstration of two qubit quantum algorithms
\cite{DiCarlo2009}, high fidelity single qubit gates \cite{Chow2010a}, high fidelity two \cite{Steffen2006,Chow2010} and
three \cite{DiCarlo2010, Neeley2010} qubit entangled states,
and Bell violation \cite{Ansmann2009}.
%Lucero2008,Chow2009,

\begin{figure}\begin{center}
\includegraphics[width=.45\textwidth]{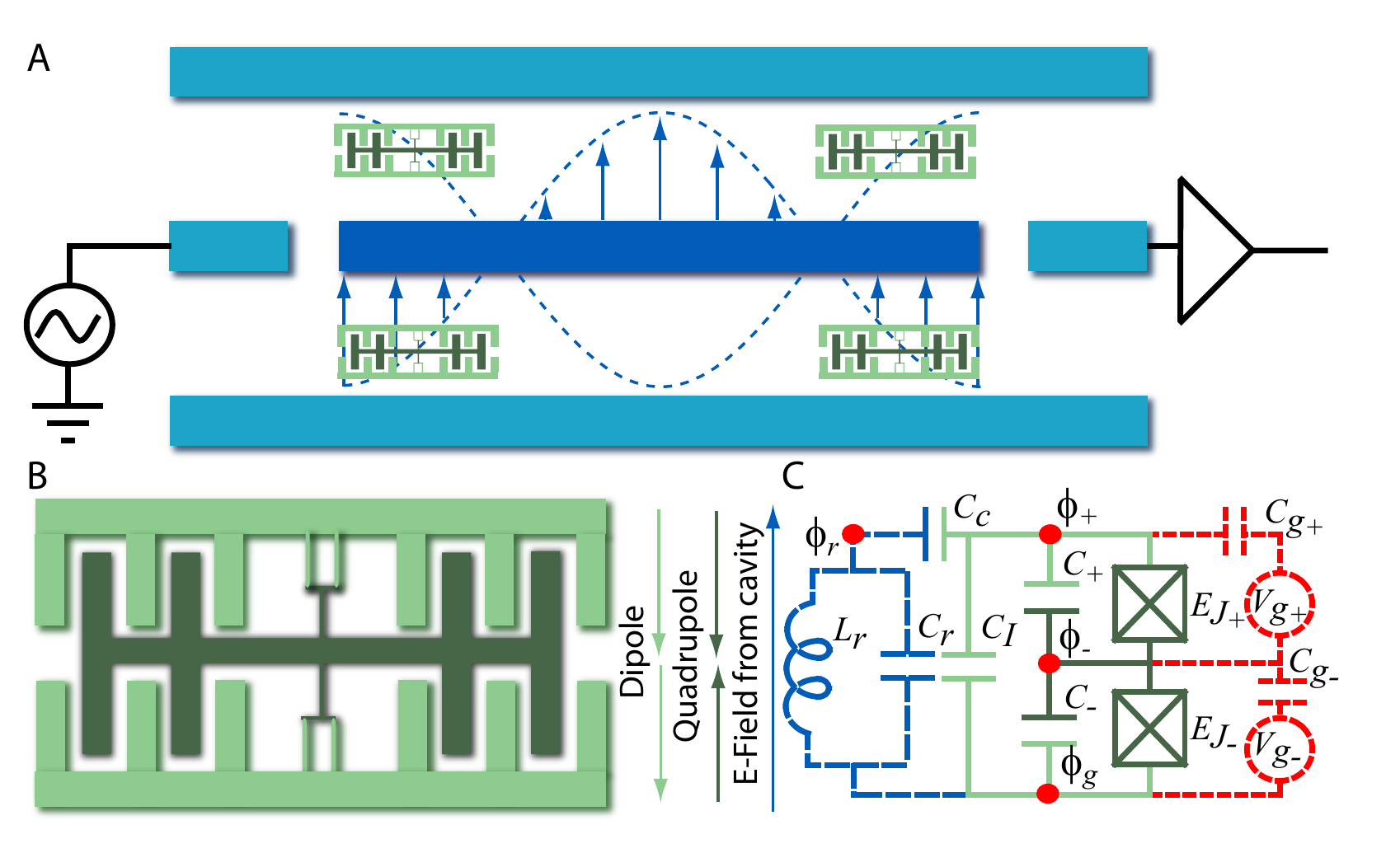}
\caption{\label{Fig:TCQ}(color online) A) is a schematic of the circuit QED architecture. B) is a schematic of the proposed three island device. The islands are connected by SQUIDs and the
arrows are used to indicated the dipole and quadrupole moments of the device. C) Is the circuit model of the device and the variables are explained within the text.}\end{center}
\end{figure}

Currently the most successful superconducting qubits are the flux \cite{Orlando1999}, phase \cite{Martinis2005}, and transmon \cite{Koch2007}
as these qubits are essentially immune to offset charge (charge noise) by design. The transmon receives its charge noise immunity by operating at a point  in parameter space where the energy level variations with offset charge are exponentially suppressed. This suppression has
experimentally been observed and
resulted in these qubits being approximately $T_1$ limited ($T_2\approx 2T_1$) in the circuit quantum electrodynamics (QED) architecture
\cite{Schreier2008}.
 In this architecture the qubits are coupled to a coplanar waveguide resonator through a Jaynes-Cummings Hamiltonian operated in the
 dispersive regime \cite{Blais2007}.
This resonator acts as the channel to control, couple, and readout the state of the qubit (see Fig. \ref{Fig:TCQ} A).

In the circuit QED architecture, a significant source of $T_1$ has been shown to be Purcell decay \cite{Houck2008}.
This is a fundamental relaxation that arises when a qubit is coupled to a resonator. It can be understood as dressing of
the qubit, the dressed qubit contains a photonic part proportional to the coupling strength $g$
divided by the detuning $\Delta$ between the qubit and resonator. This
photonic part of the qubit will relax at a rate given by the decay rate of the resonator $\kappa$ and as a result the dressed
qubit will relax at a rate given $\gamma_1^\mathrm{pur}=\kappa g^2/\Delta^2$. To overcome the Purcell decay we can either work at large detunings,
use a Purcell filter \cite{Reed2010}, or design a Purcell protected qubit.

In this letter we will present a three island device that has the properties of a qubit (two levels, arbitrary control and measurement) and has the ability to \emph{independently} tune both the resonance frequency and coupling strength $g$, whilst still exhibiting exponential suppression of the charge noise and maintaining an anharmonicity equivalent to that of the transmon and phase qubit. This tunable coupling qubit (TCQ) can be tuned from a configuration which is totally Purcell protected from the resonator $g=0$ (in a DFS) to a position which couples strongly to the resonator with values comparable to those realized for the transmon. Furthermore, we show that in the DFS position a strong measurement can be performed. The TCQ only needs to be moved from the DFS position when single and two qubit gates are required and as such in the off position all multi-qubit coupling rates are zero. That is, the TCQ in the circuit QED architecture (see Fig. \ref{Fig:TCQ} A) is its own tunable coupler to any other TCQ, going beyond the nearest neighbour tunable couplers presented in Refs. \cite{Hime2006} and \cite{Bialczak2010}.

The essential idea behind the TCQ is a three island version of the transmon as shown in Fig. \ref{Fig:TCQ} B. This device, like the transmon, has a dipole moment between each island. These dipoles can add in parallel, resulting in a larger dipole moment, or in antiparallel, creating a quadrupole moment. This device now supports two different modes and with only the dipole moment being able to couple to the resonator. That is, the mode corresponding to the quadrupole moment is a DFS with respect to Purcell decay and can be used for the storing of quantum information. Due to the capacitance between the top and lower island $C_I$ these modes couple and the ratio of quadrupole to dipole moment can be tuned by changing the energy of the upper and lower island.

The reduced circuit model we use for this device is shown in Fig. \ref{Fig:TCQ} C. The solid green lines represent
  the components associated with the TCQ with light indicating the upper $``+"$ and lower island $``-"$
 and the dark representing the center island. Each island is connected by a Josephson junction of energy $E_{J_\pm}$ and capacitance $C_\pm$ which is taken to include the Josephson capacitance. The resonator is approximated by a parallel LC circuit (blue dashed lines) with inductance
 $L_r$ and capacitance $C_r$. $C_{c}$ represents the capacitive interaction between the TCQ and resonator. Finally the dotted red lines represent charge noise resulting from voltage fluctuations $V_{g_\pm}$  that occur inside the device ($C_{g_\pm}$ represent the capacitor coupling for these fluctuations). Note a similar system was presented in Ref.~\cite{Rebic2009} for observation of giant non-linear Kerr effects in circuit QED. 

\begin{figure}\begin{center}
\includegraphics[width=.45\textwidth]{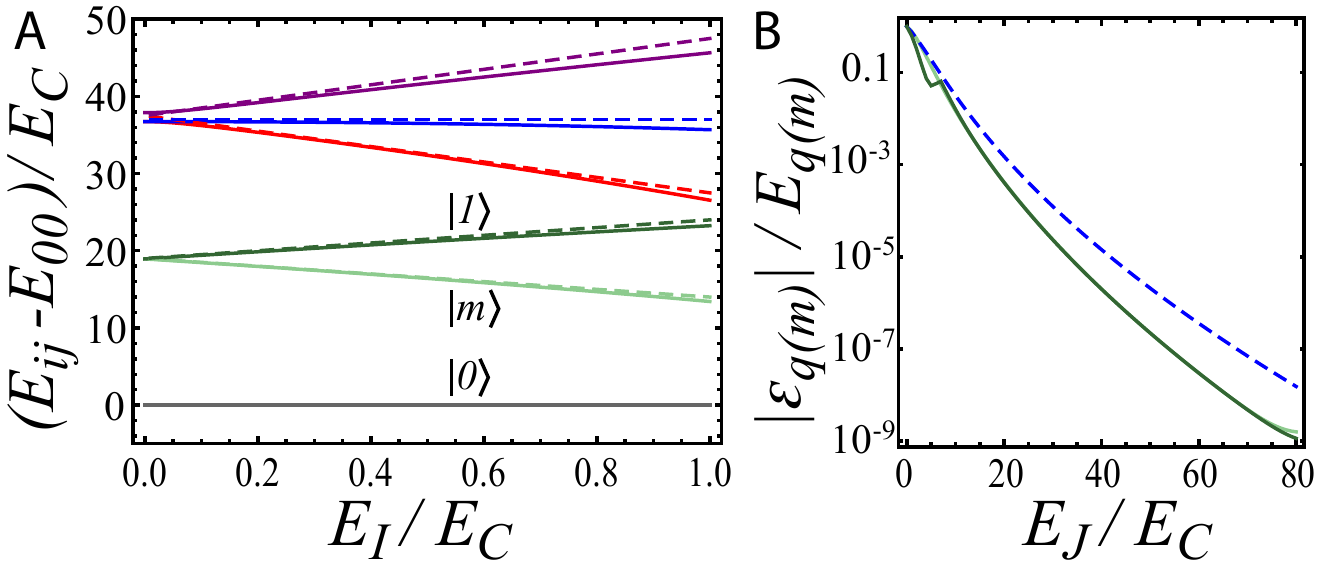}
\caption{\label{Fig:Levels}(color online) A) Eigenenergies of the TCQ Hamiltonian as a function of $E_I/E_C$ for $E_{J_\pm}=50 E_C$. Solid lines are from a numerical diagonalization and dashed lines are from the coupled anharmonic oscillator model. B) Charge dispersion $|\varepsilon_{q(m)}|$ as a function of the ratio $E_J/E_C$ for $E_I=-E_C$ (solid lines) and $E_I=0$ (dashed lines).
  }\end{center}
\end{figure}

The Hamiltonian is obtained using the method outlined in Ref. \cite{Devoret1997} with the three degrees of freedom being the phase
  across the junctions $\gamma_+=(\phi_+-\phi_-)/\Phi_0$, $\gamma_-=2\pi\phi_-/\Phi_0$, and the flux across the resonator $\phi_r$. Here $\Phi_0=h/2e$ is the flux quantum and $\phi_{\pm}$ are the node fluxes defined in Fig. \ref{Fig:TCQ} C.
   %Note $\phi_{g}=0$ as it was chosen to be the ground node. 
   We find $H=H_\mathrm{T}+H_\mathrm{I}+H_\mathrm{R}$ where $H_\mathrm{R}=\hbar\omega_r a^\dagger a$ is the Hamiltonian of the resonator with $\omega_r= 1/\sqrt{L_r C_r'}$ and $a$ being the standard annihilation operator, $H_\mathrm{T}$ the Hamiltonian for the TCQ given by
 \begin{equation}\label{eq:TCQ}
 H_\mathrm{T}=\sum_\pm 4 E_{C_\pm} (n_\pm-n'_{g_\pm})^2 -\sum_\pm  E_{J_\pm}\cos(\gamma_\pm) + 4 E_I n_+ n_-
 \end{equation} with charging energy $E_{C_\pm}=e^2/2  C'_\pm$, interaction energy $E_I=e^2/{C}'_I$, and dimensionless gate voltage $n'_{g_{\pm}}=n_{g_{\pm}}+n_{g_{\mp}}  C'_{\mp}/ C'_I$ where $n_{g_\pm}=C_{g_\pm}V_{g_\pm}/2e$. Note the prime above the capacitance indicate that they have been renormalized by the interactions. Finally $H_\mathrm{I}$ represents the interaction of the resonator with the TCQ and is
\begin{equation}\label{eq:int}
H_\mathrm{I} = 2e^2 V_\mathrm{rms}\left(\beta_+{n_+}+\beta_-{n_-}\right) (-i a^\dagger +i a)
\end{equation}  where $\beta_\pm=C_c C_{\Sigma_\mp}/[C_{\Sigma_+}C_{\Sigma_-}+(C_I+C_c)(C_{\Sigma_+}+C_{\Sigma_-})]$ with $C_{\Sigma_\pm}=C_\pm+C_{g_\pm}+C_c$, and $V_\mathrm{rms}=\sqrt{\hbar\omega_r/2 C'_r}$.  In the limit where the TCQ is isolated ($C_c=0$) and symmetric (drop all $\pm$ dependence in capacitors) then $E_{C_+}=E_{C_-}=E_C$ with $E_C=e^2(C_I+C_\Sigma)/2(C_\Sigma^2+2C_I C_\Sigma)$ and $E_I=-2E_C C_I/(C_I+C_\Sigma)$. $E_I$ can be tuned from zero to $-2E_C$ by modifying $C_I$, governed by the direct capacitance between the upper and lower island. This can be made much larger then the interaction energy between two superconducting qubits that are coupled virtually by a resonator \cite{Blais2007}. The eigenenergies of the TCQ Hamiltonian are shown in Fig. \ref{Fig:Levels} A (solid lines) as a function of $E_I/E_C$ for $E_{J_\pm}=50 E_C$. The system has a ``V'' like structure with two levels in the first excitation manifold and three in the next. As we increase $E_I$ the degeneracies in the manifolds are lifted and we have a multilevel atom with non ladder like structure.

In the limit of large $E_{J_\pm}/E_{C_\pm}$ the values of $\gamma_\pm$ are restricted to be around zero; thus we can expand the cosine terms up to fourth order \cite{Koch2007}. That is, we can model the system as two coupled anharmonic oscillators with the Hamiltonian $H_\mathrm{T_{eff}} = H_++H_-+H_{c}$ with $H_\pm=\hbar[\omega_\pm +\delta_\pm (b_\pm^\dagger b_\pm-1)/2]b_\pm^\dagger b_\pm$ and $H_{c}=\hbar J(b_+b_-^\dagger +b_+^\dagger b_-)$ with $\omega_\pm=\sqrt{8 E_{J_\pm}E_{C_\pm}}/\hbar-E_{C_\pm}/\hbar$, $\delta_\pm=-E_{C_\pm}/\hbar$, and $J=E_I (E_{J_+}E_{J_-}/E_{C_+}E_{C_-})^{1/4}/\sqrt{2}\hbar$. This Hamiltonian can be approximately diagonalized (in orders of $\delta_\pm/(\tilde\omega_+-\tilde\omega_-)$ with the transformation $D=\exp[\lambda(b_+ b_-^\dagger-b_+^\dagger b_-)]$ with $\lambda=\tan^{-1}(2J/\eta)/2+\theta$ and $\eta=\omega_+-\omega_--(\delta_+-\delta_-)/2$. Here $\theta=0$ for ($J,\eta)>0$, $\theta=\pi/2$ for $\eta<0$, and $\theta=\pi$ for $(-J,\eta)>0$. Doing this we find
\begin{equation}
\label{Ed:TIDeff}
\tilde H_\mathrm{T_{eff}}= \hbar\sum_\pm[\tilde\omega_\pm+\frac{\tilde\delta_\pm}{2}( \tilde b_\pm^\dagger \tilde b_\pm-1)] \tilde b_\pm^\dagger  \tilde b_\pm+\tilde\delta_{c}  \tilde b_+^\dagger \tilde b_+ \tilde b_-^\dagger b_-,
\end{equation}
where $\tilde \omega_\pm =\omega_\pm +(\tilde{\delta}_\pm-\delta_\pm)/2+(\delta_++\delta_-)J^2/2\mu^2 \pm \mu/2\mp\eta/2$, $\tilde\delta_{\pm}=(\delta_++\delta_-)(1+\eta^2/\mu^2)/4\pm\eta(\delta_+-\delta_-)/2\mu$ and $\tilde\delta_{c}= 2J^2(\delta_++\delta_-)/\mu^2$ with $\mu=\sqrt{4J^2+\eta^2}$ and the tilde indicating the diagonalized frame. The coupling has induced a conditional anharmonicity $\tilde\delta_{c}$, it is this anharmonicity that makes this system different to two coupled qubits, it ensures that $E_{11}$ is not equal to $E_{01}+E_{10}$. Here we have introduced the notation that superscript ${ij}$ refers to $i$ excitations in the dark mode ($``+"$) and $j$  excitations in the bright mode ($``-"$). The choice of these names will become clearer latter. The dotted lines in Fig. \ref{Fig:Levels} A are the predictions from this effective model, which agree well with the full numerics. Thus from the effective model the anharmonicities are all around $E_C$ provided $|J|>|\eta|$. That is the TCQ has not lost any anharmonicity in comparison to the transmon or phase qubit and with simple pulse shaping techniques arbitrary control of the lowest three levels will be possible \cite{Motzoi2009}. We will now introduce the notation that the qubit is formed by the space $\ket{0}=\ket{\tilde {00}}$, $\ket{1}=\ket{\tilde {10}}$ and $\ket{m}=\ket{\tilde {01}}$ is the measurement state.

Since charge fluctuations are one of the leading sources of noise in superconducting circuits we want to ensure that quantum information in the TCQ is not destroyed by charge noise. Following Ref.~\cite{Koch2007}, the dephasing time $T_\phi$ for the qubit and $m$ level will scale as $1/|\varepsilon_{q(m)}|$ where $\varepsilon_{q(m)}$ is the the peak to peak value for the charge dispersion of the $0-1$ and $0-m$ transition respectively. The dispersion in the energy levels arises from the gate charges $ n'_{g_\alpha}$ and the fact the the potential is periodic.  This can not be predicted with the coupled anharmonic oscillator model and as such is investigated numerically. We expect, that like the transmon, this will exponentially decrease with the ratio of $E_J/E_C$ as in this limit the effects of tunneling from one minima to the next becomes exponentially suppressed. This is confirmed in Fig. \ref{Fig:Levels} B where the have plotted $|\varepsilon_{q(m)}|/E_{q(m)}$ (the numerical maximum and minimum of the energy level over $ n'_{g_\alpha}$)  as a function of $E_J/E_C$ for $E_I=E_C$ and $E_I=0$ (transmon limit). 
%As expected $|\varepsilon_{q(m)}|$ reduces exponentially with the ratio $E_J/E_C$ for both states.  
That is, the TCQ has the same charge noise immunity as the transmon.

We now investigate how the two modes of the TCQ couple to the resonator. We start by applying the anharmonic oscillator model to Eq. \eqref{eq:int}. Doing this we find that $\tilde H_\mathrm{I_{eff}}= \hbar\sum_\pm \tilde g_\pm(a \tilde b_\pm^\dagger  + a^\dagger \tilde b_\pm )$ where
$\tilde g_\pm=g_\pm \cos(\lambda) \pm g_\mp \sin(\lambda)$ with $g_\pm=\sqrt{2}e^2\beta_\pm V_\mathrm{rms}(E_{J_\pm}/8E_{C_\pm})^{1/4}/\hbar$. Thus one of these coupling strengths can be set to zero for an appropriately chosen $\lambda$. For the case when $J$ is negative it is the $\tilde g_+$ rates which can be set to zero, hence the name dark for the $``+"$ and bright for the $``-"$ state. We have numerically confirmed that this model approximately predicts the matrix elements (coupling rates) for the first two manifolds (six levels). Using these six levels we can use perturbation theory to find an effective description of the situation where both the bright and dark modes of the TCQ have a dispersive interaction with the resonator, $|\Delta_\pm|\gg |\tilde g_\pm|$ with $\Delta_\pm=\tilde\omega_\pm-\omega_r$. To second order in $\tilde g_\pm /\Delta_\pm$ we find that the resonator-TCQ interaction induces both a Lamb shift on the TCQ and a TCQ state dependent pull on the resonator. The qubit ($0-1$) transition frequency becomes $\omega_q=\tilde\omega_++\tilde g^2_+/\Delta_+$ and the $0-m$ transition frequency becomes $\omega_m=\tilde\omega_-+\tilde g^2_-/\Delta_-$. The resonator frequency is $\omega_r+\chi_{k}$ with resonator pull $\chi_{0}=-\tilde g_-^2/\Delta_--\tilde g_+^2/\Delta_+$, $\chi_{1}=(\tilde\delta_+-\Delta_+)\tilde g_+^2/\Delta_+(\tilde\delta_++\Delta_+)-\tilde g_-^2/(\tilde \delta_c+\Delta_-)$, and $\chi_m=(\tilde\delta_--\Delta_-)\tilde g_-^2/\Delta_-(\tilde\delta_-+\Delta_-)-\tilde g_+^2/(\tilde \delta_c+\Delta_+)$. All states $|0\rangle$, $|1\rangle$,  and $|m\rangle$ have a different resonator frequency and thus can be measured by probing the transmission of the resonator.

To calculate the induced decay rate on the qubit by relaxation of the resonator (Purcell decay) we can use the same perturbation theory. This amounts to evaluating $\gamma_1^\mathrm{pur}=\kappa|\langle\bar 1|a|\bar0\rangle|^2$ where the bar is the first order correction to the TCQ levels from the resonator-TCQ interaction. Doing this we find  $\gamma_p=\kappa \tilde g_+^2/\Delta_+^2$. Thus, setting $g_+$ to zero the Purcell effect (to second order) will be canceled and yet the difference between the resonator pull for the $\ket{0}$ and $\ket{1}$ state is non-zero and given by $\chi=\tilde \delta_c \tilde g_-^2/\Delta_-(\Delta_-+\tilde\delta_c)$. This has the same functional form as the transmon (when $|J|>|\eta|$) and is detectable with current microwave electronics.  Furthermore, signal-to-noise (SNR) arguments from Ref.~\cite{Gambetta2008} carry over to this system and we find $\mathrm{SNR}=4 n \eta\chi T_m$. Here $\eta$ is efficiency of collecting the photons emitted from the resonator, $n$ is the number of the photons in the resonator which should not exceed $n_\mathrm{crit}=\Delta_-^2/4\tilde g_-^2$ \cite{Gambetta2008}, and $T_m$ is the measurement time. Taking realistic values $T_m=1 \mu$s, $\chi/2\pi=10$ MHz, and $\eta=1/20$ gives a SNR around $13 n$.  However, much higher values can be obtained by using the following protocol: Set $\omega_m=\omega_r$ and have $C_c$ large enough to ensure that $|\omega_q-\omega_r|\gg |\tilde g_\pm|$ then the $0-m$ transition will vacuum Rabi split the cavity transmission if the qubit is in the 0 state otherwise the transmission will be the bare resonator. That is, the heterodyne power $\kappa|\langle a\rangle|^2$ in steady state will be $\kappa\xi^2(\tilde g_-^2+\kappa^2/16)/(\tilde g_-^2+\kappa^2/16+\xi^2)^2$ \cite{Bishop2009} for the $\ket{0}$ and $4\xi^2/\kappa$ for the $\ket{1}$ state with $\xi$ being the cavity drive amplitude at drive frequency $\omega_r$. Taking $\mathrm{SNR}=\eta T_m\kappa(|\langle a\rangle_0|^2-|\langle a\rangle_1|^2)$ we find values as large as 1000 for $\tilde g_-/2\pi=100$ MHz, $\kappa/2\pi= 10$ MHz, and $\xi=\tilde g_-$. This is similar to a cycling-type measurement and is quantum non-demolition (after wait time $\kappa$).  There will be limitations impose on size of $\xi$ due to finite anharmonicity of the TCQ.

\begin{figure}\begin{center}
\includegraphics[width=.45\textwidth]{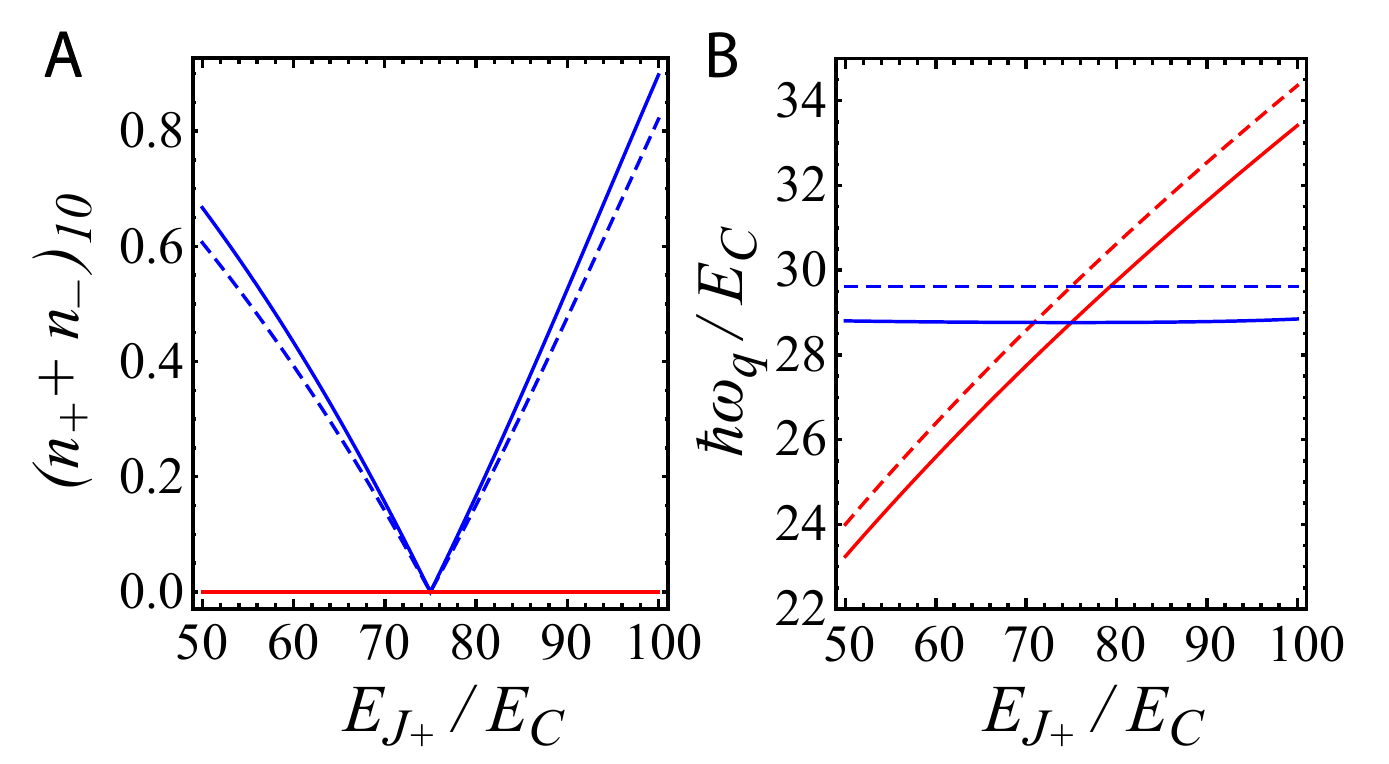}
\caption{\label{Fig:Tune}(color online) Matrix element of the collective Cooper pair number operator (A) and transition energy (B) of the dark state as a function of the energy ratio $E_{J_+}/E_C$ for $ n'_{g_+}= n'_{g_-}=0$, $E_I=-E_C$ and $E_{J_-}$ is numerically solved to ensure that only the coupling strength is tuned (blue) and frequency (red). Solid lines are from a numerical diagonalization and dashed lines are from the coupled anharmonic oscillator model.
  }\end{center}
\end{figure}

To achieve tuning of $\tilde g_+$ we modify the original circuit and replace the Josephson junctions by SQUIDs with Josephson energy $E_{J_\pm}^{(1)}$ and $E_{J_\pm}^{(2)}$ (this is hinted at in Fig. \ref{Fig:TCQ} B). In making this replacement the only change in the above theory is the replacement $E_{J_\pm}\rightarrow E^\mathrm{max}_{J_\pm}\cos(\pi\Phi_{x_\pm}/\Phi_0)\sqrt{1+d^2\tan^2(\pi\Phi_x/\Phi_0)}$ with $E^\mathrm{max}_{J_\pm}=E_{J_\pm}^{(1)}+E_{J_\pm}^{(2)}$, $d=(E_{J_\pm}^{(1)}-E_{J_\pm}^{(2)})/E^\mathrm{max}_{J_\pm}$ and $\Phi_{x_\pm}$ is the external flux applied to each SQUID which we assume to be independent (this is not required but simplifies our argument). This independent control allows us to change $E_{J_\pm}$ independently which in-turn allows independent control on $\tilde g_+$ and $\omega_q$. To illustrate this we consider the symmetric case and
plot in Fig. \ref{Fig:Tune} the normalized coupling strength $\tilde g_+\hbar/2e^2 V_\mathrm{rms} \beta$ (A) and $\tilde\omega_q$ (B) as a function of the ratio $E_{J_+}/E_C$ when $E_I=-E_C$ and $E_{J_-}$ is numerically solved to ensure that only the coupling rate (blue) and frequency (red) vary respectively for both the full numerical (solid) and effective model (dashed). In the full numerical model $\tilde g_+= 2e^2 V_\mathrm{rms}\left\bra{1}(\beta_+{n_+}+\beta_-{n_-}\right) \ket{0}/\hbar$. Here the independent control is clearly observed.  Note that while our numerical investigation was only for the symmetric case independent tunable $\tilde g_+$ (from zero to large values) and $\omega_q$ will still occur when the device is not symmetric. There is just a different condition on $E_{J_\pm}$ for the required tuning.

With the extra control channel there is the possibility of additional qubit decoherence from flux fluctuations. A reasonable estimate for the flux induced relaxation is $\gamma^\mathrm{flux}_1=\sum_\pm|\langle{0}| \partial H/\partial \Phi_{x_\pm}  |1\rangle|^2 M^2_\pm S_{I_\pm}(\omega_q)/\hbar^2$ where $M_\pm$ is the mutual inductance between the bias line and the TCQ, and $S_{I_\pm}(\omega_q)$ is the current noise in the bias line which at low temperatures $S_{I_\pm}(\omega_q)\approx \hbar \omega_q/R$  \cite{Schoelkopf2003}. Taking $M=200~\Phi_0/A$, $R=50~\Omega$, $E^\mathrm{max}_{J_\pm}/h=20$ GHz, $E_C/h=E_I/h=0.35$ GHz, and $d=10\%$ we find $T_1\approx 1$s. To estimate the contribution to dephasing we assume the noise is $1/f$ and from Ref. \cite{Martinis2003} the dephasing time $T_\phi\approx|\partial \omega_q/\partial \Phi_{x_\pm}|^{-1}/A_\phi $ where $A_\phi$ is the flux noise measured at 1 Hz which for similar superconducting devices has been measured to be $10^{-6}\Phi_0/\sqrt{\mathrm{Hz}}$ \cite{Yoshihara2006}. This predicts a  $T_\phi\approx 20\mu$s (at $\Phi_{x_\pm}=\pi\Phi_0/4$). This rates are consistent with the $T_1$ and $T_\phi$ predictions for the transmon (with only frequency control) from flux noise \cite{Koch2007}. That is, in comparison to the transmon the additional flux line that gives both independent control of the coupling and frequency has not added any extra noise.

In conclusion we have presented a new device for quantum information processing with superconducting circuits. It is an extension of the transmon and uses quantum interference to achieve independent control of both the coupling strength and frequency. Furthermore it can be tuned to a configuration where it is in a decoherence free subspace with respect to relaxation induced by the Purcell effect whilst still allowing efficient readout of its quantum state. It also offers the possibility of implementing a cycling-type measurement. 

\begin{acknowledgments}
We acknowledge D. I. Schuster, L. S. Bishop, and R. J. Schoelkopf for valuable discussions. JMG was supported by CIFAR, Industry Canada, MITACS, MRI and NSERC. A. B. by NSERC, the Alfred P. Sloan Foundation and CIFAR. AAH by the Alfred P. Sloan Foundation and the David and Lucile Packard Foundation.
\end{acknowledgments}

%\bibliography{/home/jgambett/Subversion/private/References/Reference}
%\bibliographystyle{/home/jgambett/Subversion/private/References/mybibtexstylePRL}

\end{document}